\documentclass[aps,twocolumn,epsf,floats,prl,footinbib]{revtex4-1}
\usepackage{graphics,graphicx,epsfig}
\usepackage{amssymb,color}
\usepackage{epsf,epstopdf,wrapfig}
\usepackage{amsmath}
\usepackage{amsmath}
\usepackage{amssymb}
\usepackage{graphicx}
\usepackage{bm}
\usepackage{xr}
\usepackage{scrextend}

\def\(({\left(}
\def\)){\right)}                       
\def\[[{\left[}
\def\]]{\right]}

\def\R0{{\cal R}_0}
\def\Rl{{\cal R}_l}

\newcommand{\beq}{\begin{equation}}
\newcommand{\eeq}{\end{equation}}
\newcommand{\bea}{\begin{eqnarray}}
\newcommand{\eea}{\end{eqnarray}}

\begin{document}

\title{Dynamical renormalization group approach to the collective behaviour of swarms}

\author{
Andrea Cavagna$^{1,2}$, 
Luca Di Carlo$^{2,1}$,
Irene  Giardina$^{2,1,3}$, 
Luca Grandinetti$^{4}$,
Tomas S. Grigera$^{5,6,7}$,
Giulia Pisegna$^{2,1}$
}

\affiliation{$^1$ Istituto Sistemi Complessi, Consiglio Nazionale delle Ricerche, UOS Sapienza, 00185 Rome, Italy}
\affiliation{$^2$ Dipartimento di Fisica, Universit\`a\ Sapienza, 00185 Rome, Italy}
\affiliation{$^3$ INFN, Unit\`a di Roma 1, 00185 Rome, Italy}
\affiliation{$^4$ Dipartimento di Scienza Applicata e Tecnologia, Politecnico di Torino, Torino, Italy}
\affiliation{$^5$ Instituto de F\'\i{}sica de L\'\i{}quidos y Sistemas Biol\'ogicos CONICET -  Universidad Nacional de La Plata,  La Plata, Argentina}
\affiliation{$^6$ CCT CONICET La Plata, Consejo Nacional de Investigaciones Cient\'\i{}ficas y T\'ecnicas, Argentina}
\affiliation{$^7$ Departamento de F\'\i{}sica, Facultad de Ciencias Exactas, Universidad Nacional de La Plata, Argentina}

\begin{abstract}
We study the critical behaviour of a model with non-dissipative couplings aimed at describing the collective behaviour of natural swarms, using the dynamical renormalization group.  At one loop, we find a crossover between a conservative yet unstable fixed point, characterized by a dynamical critical exponent $z=d/2$, and a dissipative stable fixed point with $z=2$, a result we confirm through numerical simulations. The crossover is regulated by a conservation length scale that is larger the smaller the effective friction, so that in finite-size biological systems with low dissipation, dynamics is ruled by the conservative fixed point.  In three dimensions this mechanism gives $z=3/2$, a value significantly closer to the experimental result $z\approx 1$ than the value $z\approx 2$ found in fully dissipative models, either at or off equilibrium.  This result indicates that non-dissipative dynamical couplings are necessary to develop a theory of natural swarms fully consistent with experiments.
\end{abstract}

\maketitle

Collective behaviour in biological groups emerges when local interactions give rise to correlations that significantly exceed the scale of the individuals. According to this somewhat restricted and yet compelling definition, collective behaviour seems the ideal hunting ground for statistical physics, and in particular for its most powerful theoretical tool, the Renormalization Group (RG) \cite{wilson1979problems}. In statistical physics, a taxing but crucial requirement a successful theory of a collective phenomenon must meet is to reproduce the right form of the correlation functions and, most importantly, the correct values of the critical exponents \cite{patashinskii_book}, the calculation of which is the RG task. It is not surprising, then, that the RG can be applied to collective biological systems; the hydrodynamic theory of flocking of Toner and Tu is a pioneering step in this direction \cite{toner1998flocks}. Here we adopt an RG approach to the collective dynamics of swarms.

Experiments on large natural swarms in the field  \cite{cavagna2017dynamic}, show two things: {\it i)} dynamic correlations of the velocities have an inertial form incompatible with the classic exponential relaxation of overdamped systems, and {\it ii)} critical slowing down, namely the relation linking relaxation time and correlation length, $\tau \sim \xi^z$, holds with a dynamical critical exponent $z\approx 1$, very unusual for purely dissipative dynamics. Both facts urge for a theoretical explanation.  The simplest, and yet most far-reaching model of collective behaviour in biological systems was introduced by Vicsek and co-workers \cite{vicsek+al_95}: it describes individuals as self-propelled particles moving at fixed speed and aligning their velocities to those of their neighbours through a so-called `social force' \cite{vicsek_review, marchetti_review}.
Vicsek's model is analogous to a ferromagnetic system (velocities playing the role of local magnetizations) with fully dissipative Langevin dynamics; however, at variance with equilibrium ferromagnets, the interaction network in the Vicsek model changes in time, due to the self-propulsion of the individuals.  
In its great flexibility, the Vicsek model describes both polarized flocks and unpolarized swarms, depending on noise and density; therefore, in \cite{cavagna2017dynamic}, dynamical correlations in the swarm phase of the Vicsek model were measured. In contrast with real experiments, though, Vicsek swarms display purely exponential relaxation and a dynamical critical exponent $z\approx 2$. The fact that self-propelled Vicsek swarms have the same critical exponent as equilibrium dissipative models 
\footnote{At least up to sizes $N=10^3$, comparable with experimental swarms, which is the regime studied in \cite{cavagna2017dynamic}}, suggests that out-of-equilibrium  effects due to self-propulsion may not be the primary cause of anomalous relaxation in natural swarms. Hence, we focus here on developing a theory with fixed interaction network, but with a novel type of dynamical coupling, leaving the self-propelled generalization to future studies.

An inertial form of the dynamic correlation function suggests  that the equations of motion contain conservative terms deriving from a symmetry of the interaction \cite{hohenberg1977theory}. When this happens, the force between the individuals is mediated by the symmetry generator, which is a conserved momentum conjugate to the primary degree of freedom through some inertia \cite{cavagna2018physics}. The Inertial Spin Model (ISM), introduced in \cite{attanasi+al_14,cavagna+al_15} for the description of information transfer in flocks, contains a non-dissipative coupling between the velocities and the generator of rotations, called {\it spin}. It was therefore suggested in \cite{cavagna2017dynamic} that the ISM may also describe the inertial form of swarm relaxation. Moreover, non-dissipative couplings are known \cite{hohenberg1977theory} to lower $z$ below its purely dissipative value $\approx2$. Therefore, it seems possible that the ISM may help with both experimental traits of swarm dynamics. 
From another point of view, though, this may seem an ill-founded hope. In Hamiltonian systems the spin is strictly conserved \cite{hohenberg1977theory}, while in biological groups it {\it cannot} be: the spin is what causes animals to turn, hence it must be dissipated in absence of interaction or perturbation \cite{cavagna2018physics}. Therefore, the ISM contains both non-dissipative couplings {\it and} effective friction. Because friction always takes over in the hydrodynamic limit,  one may expect the ISM to have the same critical dynamics as an over-damped system. Here we resolve this quandary by studying the critical dynamics of the fixed-network ISM through a renormalization group approach. A full account of our calculation can be found in \cite{companion}.

The field-theory ISM dynamics in the fixed-network case is defined by the following equations,
\begin{align}
\frac{\partial \bm\psi}{\partial t} &= -\Gamma_0 \frac{\delta \cal H}{\delta \bm\psi} + g_0\, \bm\psi \times \frac{\delta \cal H}{\delta \bm s} + \bm\theta  \ ,
\label{ISM1} \\
\frac{\partial \bm s}{\partial t} &= -\left(\eta_0-\lambda_0\nabla^2\right) \frac{\delta \cal H}{\delta \bm s} + g_0 \, \bm\psi\times\frac{\delta \cal H}{\delta\bm\psi} + \bm \zeta \ ,
\label{ISM2}
\end{align}
where the coarse-grained field $\bm\psi(x,t)$ is a vectorial order parameter, corresponding to the velocity in the self-propelled case, while the conjugated momentum
$\bm s(x,t)$ represents the coarse-grained spin, namely the generator of the rotational symmetry acting upon $\bm\psi$ (i.e.\ the Poisson bracket reads, $\{\psi_\mu,s_\nu\}=g_0\epsilon_{\mu\nu\rho}\psi_\rho$, where $\epsilon_{\mu\nu\rho}$ is the Levi-Civita antisymmetric symbol \footnote{More precisely, the field {\it canonically} conjugate to $\bm s$ is the phase $\varphi$ of the primary field $\bm\psi$, not $\bm\psi$ itself; this is the reason why cross products enter the dynamical equations -- see \cite{cavagna2018physics}.}). The white Gaussian noises $\bm\theta$ and $\bm\zeta$ have variance $2\Gamma_0$ and $2(\eta_0 +\lambda_0 k^2)$, respectively. The effective Hamiltonian has the standard rotationally symmetric form \cite{hohenberg1977theory},
\beq
{\cal H} = \int d^dx \ \left\{ \frac{1}{2}(\nabla \bm\psi)^2 + r_0 \, \psi^2 + u_0 \psi^4 + \frac{s^2}{2\chi_0} \right\} \ ,
\eeq
where the square gradient enforces the alignment interaction, and instead of the fixed speed constraint, $\lvert\bm\psi\rvert^2=1$, of the microscopic model, one has the confining potential, $r_0 \psi^2 + u_0 \psi^4 $, where $r_0<0$ in the ordered phase. The parameter $\chi_0$ is the effective inertia associated to the spin; at the static level, $\bm s$ is a Gaussian field, hence there are no corrections to the naive scaling dimension of $\chi_0$, so we can fix $\chi_0=1$ in the following. The kinetic coefficient $\Gamma_0$, the effective friction $\eta_0$, the transport coefficient $\lambda_0$ and the non-dissipative coupling constant $g_0$, are dynamical parameters. When working in Fourier space, all integrations over $k$ are performed up to a cutoff $\Lambda$, which corresponds to the inverse of a microscopic length scale, of the order of the inter-particle distance. 

Compared to the microscopic ISM \cite{cavagna+al_15}, there are additional terms in (\ref{ISM1}-\ref{ISM2}) coming from the coarse-graining. The diffusive term  $\Gamma_0\nabla^2\bm\psi$ (and its coupled noise $\bm\theta$) is sub-leading in the highly polarized (flock) phase \cite{cavagna2015silent}, but it is crucial in the near-critical phase of swarms. The non-dissipative coupling between the two fields is now ruled by a coupling constant $g_0$, to make scaling dimensions explicit. Finally, even though the spin is microscopically dissipated only through the non-conservative friction, $-\eta_0 \bm s$, the RG calculation shows that, as an effect of the symmetry, the first correction to the self-energy of $\bm s$ is of order $k^2$ \cite{companion}; this means that the coarse-graining of the microscopic spins produces a conservative transport term of the type $\lambda_0 \nabla^2 \bm s$, thus requiring that we insert this term in the field equations from the outset. For $\eta_0=0$ our model is identical to the Heisenberg antiferromagnet, and, in the planar case, to superfluid helium (respectively called models G and E in the literature \cite{halperin1976renormalization,hohenberg1977theory,de1978field}).

It is interesting to note that the combination $\sqrt{\lambda_0/\eta_0}\equiv\R0$ has the physical dimension of a length scale, which is larger the smaller the friction. Intuitively, we can guess that within the scale $\R0$ non-conservative effects are weaker than the conservative ones generated by the symmetry, while beyond $\R0$ the opposite occurs and dissipation takes over. As we shall see, the conservation length scale $\R0$ will indeed rule the critical dynamics of the model, leading to a non-trivial crossover between different critical exponents.

The renormalization group approach unfolds through two stages \cite{wilson1974renormalization,hohenberg1977theory}: {\it i)} integration of the short wavelengths details; {\it ii)} rescaling of momentum and frequency. In the first stage, one calculates the effective probability distribution of the long-wavelengths fields $\bm\psi(k,\omega)$, with $k<\Lambda/b$, where $b$ is a scaling factor, by integrating over all short-wavelengths fluctuations with $k$ in the so-called {\it momentum shell}, $\Lambda/b < k < \Lambda$. This integration produces corrections to the Gaussian terms in the dynamical equations for the long-wavelengths fields \cite{halperin1972calculation, halperin1976renormalization, hohenberg1977theory}. To calculate these corrections one writes the free propagators, which are defined as the inverse of the linear part of the equations of motions, 
\begin{align}
G_{b,\bm\psi}^{-1}({\bf k},\omega) &\ = -i\omega + \Gamma_0(k^2 + r_0)  - \Sigma_{b}({\bf k},\omega)   \ ,
\label{tangatour}
\\
 G_{b,\bm s}^{-1}({\bf k},\omega) &\ =  -i\omega + \eta_0 +\lambda_0 k^2  - \Pi_{b}({\bf k},\omega) \ ,
 \label{mengatour}
\end{align}
where the {\it self-energies} $\Sigma_b$ and $\Pi_b$ arise from the on-shell integration \cite{companion}. We see that in the limit $k\to 0$ the $\mathrm O(k^2)$ terms in the self-energies  correct $\Gamma_0$ and $\lambda_0$; on the other hand, a term of $\mathrm O(1)$ in the self-energy $\Pi_b$, would correct the effective friction, $\eta_0$; however, it is possible to show \cite{companion} that, as an effect of the symmetry, there is no such term in $\Pi_b$, so that $\eta_0$ has no correction from the RG  integration. Regarding the dynamical coupling constant $g_0$, one can show that, even though the spin is not conserved, the symmetry nevertheless protects the naive scaling dimension of $g_0$, which therefore does not pick up a perturbative correction \cite{companion}. 

After integrating over the momentum shell, the theory is left with a novel cutoff, $\Lambda/b$; hence, the second RG stage consists in rescaling the momentum $k$ in such a way to restore the original cutoff $\Lambda$. The key idea of the dynamical renormalization group \cite{hohenberg1977theory} is that, close to criticality, a rescaling of space entails a rescaling of time, regulated by the dynamical critical exponent, $z$. Hence, the momentum rescaling must be associated to a corresponding frequency rescaling, 
\beq
k \to b \, k \quad , \quad
\omega \to b^z \, \omega \ .
\label{rescaling}
\eeq
Shell integration and rescaling give rise to a new effective theory describing the physics of the system at longer length scales. By iterating $l$ times the RG step we obtain a set of recursive equations describing the {\it flow} of the parameters under successive coarse-grainings. At the critical point the correlation length $\xi$ is infinite and the theory at large distances is scale invariant, so that the fixed points of the RG flow give the effective values of the parameters ruling the system at large distances, i.e. for $k\to 0$ \cite{wilson1974renormalization}. The RG equations at $\xi=\infty$ are the following \cite{companion},
\begin{align}
\Gamma_{l+1} &= 
 \Gamma_l \ b^{z-2}  \left( 1 + \frac{2 f_l}{1+w_l} X_l \ln b\right) \ ,
 \label{tupo} \\
 \lambda_{l+1} &= 
 \lambda_l \ b^{z-2} \left( 1 + \frac{1}{2} f_l \ln b\right) \ ,
\label{tapa} \\ 
\eta_{l+1}&=\eta_l \  b^z  \ ,
\label{nano} \\
g_{l+1} &= g_l \ b^{z-d/2} \ ,
\label{tepa}
\end{align}
where we have introduced the effective running parameters, $f_l = \Lambda^{d-4} K_d\,  g_l^2/(\Gamma_l\lambda_l)$ ($K_d$ is the volume of  the $d$-dimensional unit sphere), $w_l=\Gamma_l/\lambda_l$,  ${\cal R}_l =\sqrt{\lambda_l/\eta_l}$, and the crossover factor, 
\beq
X_l = \frac{(1+w_l) ({\cal R}_l \Lambda)^2}{1+(1+w_l) ({\cal R}_l \Lambda)^2} \ .
\label{pixar}
\eeq
The powers of $b$ in the RG equations are due to the rescaling of $k$ and $\omega$, which rescales each parameter by its naive physical dimensions, whereas the $\ln b$ terms derive from the shell integration. For zero friction, $\eta_l=0$, the conservation length scale diverges, $\Rl=\infty$; this implies $X_l=1$, and the flow equations become identical to the fully conservative models \cite{halperin1976renormalization}. However, we see from \eqref{nano} that for any reasonable (i.e. positive) value of $z$, a non-zero initial value of the friction $\eta_0$ {\it grows} along the flow, i.e. dissipation becomes increasingly relevant at longer scales. To see the consequences of this fact we must write the closed set of recursive equations for the effective parameters,
\begin{align}
f_{l+1} &= f_l \ b^{\epsilon} \left[ 1-f_l\left(\frac{2  X_l}{1+w_l} + \frac{1}{2} \right) \ln b\right] \ , 
\label{nibbio} \\
w_{l+1} &= w_l \left[ 1+f_l \left(\frac{2 X_l }{1+w_l} - \frac{1}{2} \right) \ln b\right]\ , 
\label{gheppio} \\
{\cal R}_{l+1} &= {\cal R}_l \ b^{-1} \left[ 1+ \frac{1}{4} f_l \ln b\right] \ ,
\label{bombazza}
\end{align}
where $\epsilon = 4-d$ is the expansion parameter of the RG series \cite{wilson1972critical}; our calculation is performed at $\mathrm O(\epsilon)$, i.e. at one-loop level \cite{companion}. Once the fixed points of the RG equations are found, the dynamical critical exponent $z$ ruling the relaxation of the order parameter is obtained by imposing that the fixed point of the kinetic coefficient of $\bm \psi$, namely $\Gamma^*$, is finite \cite{hohenberg1977theory} (asterisks denote  fixed point values). This condition gives,
\beq
z= 2 - \frac{2f^*}{1+w^*} X^*   \ .
\eeq

\begin{figure}
\centering
\includegraphics[width=0.45 \textwidth]{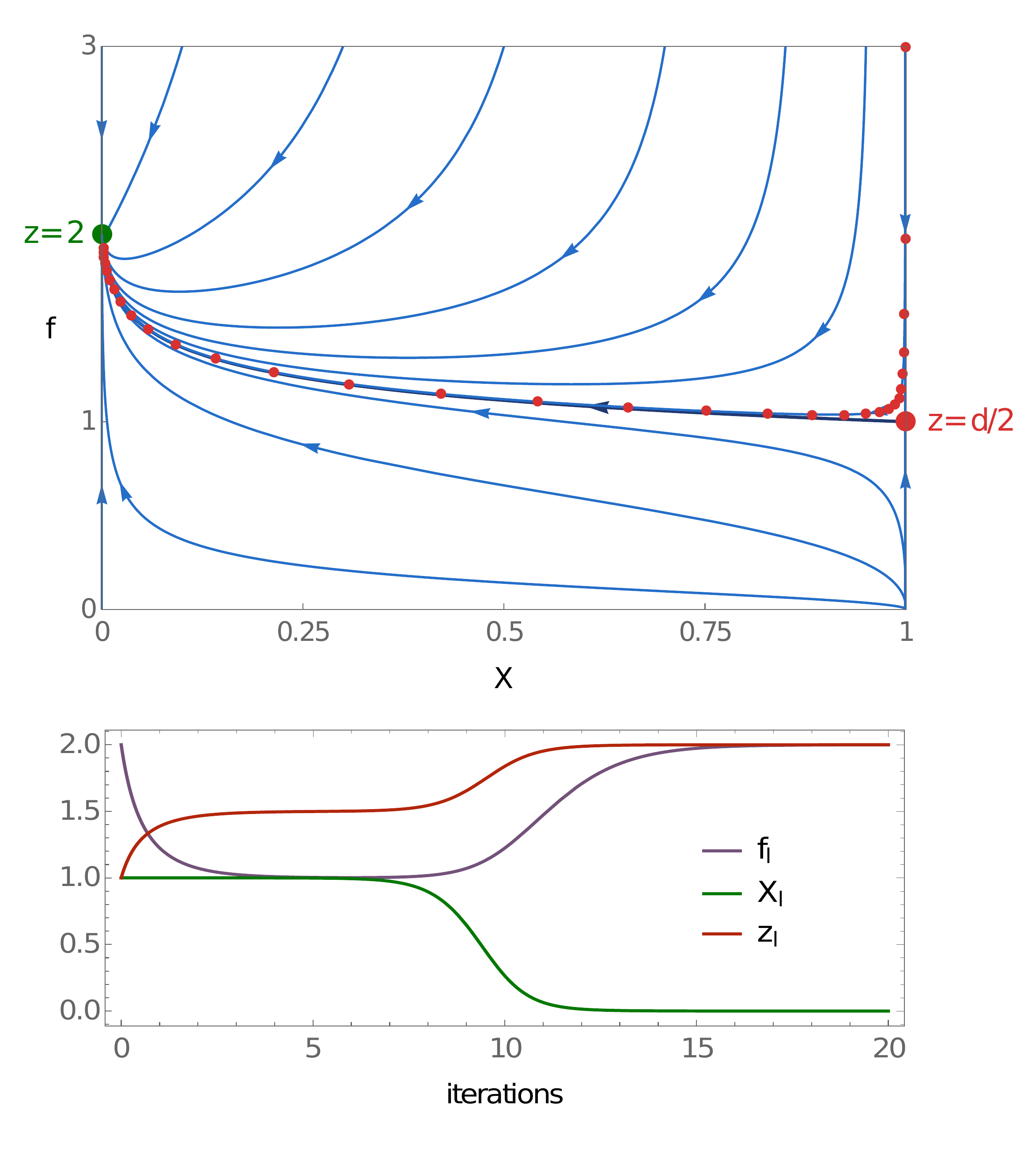}
\caption{ 
{\bf The renormalization group flow and crossover.} Top: Flow diagram on the $(X_l,f_l)$ plane for $d=3$. When the initial friction $\eta_0$ is small, i.e. $X_0\sim 1$ (red dots), the flow converges towards the unstable fixed point, $z=d/2$, and remains in its neighbourhood for many iterations, before crossing over to the $z=2$ fixed point.  Bottom: When plotting the running parameters and the critical exponent as a function of the iteration step along a flow line with small $\eta_0$, the RG crossover clearly emerges.
}
\label{fig::experiment}
\end{figure}

\noindent
The flow equations (\ref{nibbio}-\ref{bombazza}) have two fixed points; the first one is the same as the fully conservative case \cite{hohenberg1977theory}, namely,
\beq
f^* = \epsilon \ , \ w^*=3 \ , \  {\cal R}^*=\infty \ , \  X^*=1 \  \ \Rightarrow { z=d/2 }\ .
\eeq
 At this fixed point dissipation is irrelevant ($\eta^*=0$), and the conservation law entailed by the symmetry of the Hamiltonian rules the dynamics at all scales. This fixed point, though, is {\it unstable}, as it can be seen from the fact that any large, but finite, initial value of $\R0$ decreases under the RG equation \eqref{bombazza}, driving the flow to the following {\it stable} fixed point,
\beq
f^* = 2\epsilon \ , \ w^*=0 \ , \ {\cal R}^*=0 \ , \ X^*=0 \ \ \Rightarrow { z=2 }\ .
\eeq
At this fixed point the dynamics is completely taken over by dissipation ($\eta^*=\infty$); no trace remains of the non-dissipative coupling associated to the symmetry, and the conservation scale ${\cal R}^*$ shrinks to zero
\footnote{Note that $z=2$ is the Gaussian ---i.e.\ free--- value of the dynamical critical exponent; at the stable fixed point, the first corrections to $z$ are found at two-loop, $\mathrm O(\epsilon^2)$, order.}.

The interplay between the two RG fixed points gives rise to an interesting crossover, similar to what happens, at the static level, in uniaxial dipolar ferromagnets \cite{frey1990renormalized, frey1995crossover}. The flow diagram for $f_l$ and $X_l$ (Fig.~1) shows that, when the starting value of $X_0$ is close to $1$, namely when the friction $\eta_0$ in the actual model is small, the parameters rapidly converge towards the $z=d/2$ fixed point {\it and linger thereabout for many RG iterations}, before eventually crossing over to the $z=2$ fixed point. This happens because the unstable fixed point is actually {\it stable} along the line $X_0=1\, (\eta_0=0)$, thus acting as a pseudo-attractor for low-dissipation dynamics.
This RG crossover gives rise to a physical crossover in the relaxation of the system, depending on the interplay between the correlation length, $\xi$, and the conservation length, $\R0$. Both length scales decrease under the RG flow \cite{cardy1996scaling}, but while $\xi$ has scaling dimension $1$, close to the $z=d/2$ fixed point, where $f_l/4\sim \epsilon/4$, equation \eqref{bombazza} gives to $\R0$ an anomalous scaling dimension different from $1$,
\beq
\xi_{l+1} = \xi_l/b   \quad , \quad {\cal R}_{l+1}=\Rl/\, b^{d/4} \ ,
 \eeq
where the initial value of the correlation length in the RG flow coincides with its physical value, $\xi_0=\xi$. The RG iteration must stop when the correlation length has been reduced to the order of the lattice spacing \cite{cardy1996scaling}, $\xi_{l+1} \sim 1/\Lambda$, a condition equivalent to $b^l=\xi\Lambda$. In order for the flow to be still in the neighbourhood of the $z=d/2$ fixed point when the RG iteration stops, we must have $X_{l+1}\sim 1$, which, given equation \eqref{pixar}, requires ${\cal R}_{l+1} \gg 1/\Lambda$, that is $\R0/(b^l)^{d/4} \gg 1/\Lambda$. By plugging into this last relation the RG stop condition, $b^l=\xi\Lambda$, one gets that the $z=d/2$ fixed point rules for $(\xi\Lambda)^{d/4} \ll \Lambda\R0$. Conversely, for large correlation lengths the system will be ruled by the stable fixed point, $z=2$. In this way we obtain the following dynamical crossover, 
\begin{align}
\xi &\ll \R0 ^{4/d} \quad  \Rightarrow \quad \tau \sim \xi^{d/2}  \ ,
\label{cross1}
\\
\xi &\gg \R0 ^{4/d} \quad  \Rightarrow \quad \tau \sim \xi^{2}  \ ,
\label{cross2}
\end{align}
where we have set the cut-off scale to $1$ to simplify the notation. Critical slowing down is therefore governed by two different exponents, depending on the scale of the correlation, or, equivalently, on how close we are to the critical point. This result holds for finite $\xi$ at $k=0$, but an identical crossover occurs at $\xi=\infty$ when varying the relaxation scale $k$  \cite{companion}, namely, 
\begin{align}
k &\gg  \R0 ^{-4/d} \quad  \Rightarrow \quad  \tau\sim k^{-d/2}  \ ,
\label{cross11}
\\
k &\ll  \R0 ^{-4/d} \quad  \Rightarrow \quad \tau\sim k^{-2}  \ .
\label{cross22}
\end{align}
Once again, we remark that the crossover is regulated by the conservation length scale, which in turns depends on dissipation, $\R0=\sqrt{\lambda_0/\eta_0}$: the smaller the effective friction, $\eta_0$, the larger the scale up to which critical dynamics is ruled by the conservative exponent, $z=d/2$.

\begin{figure}
\centering
\includegraphics[width=0.50 \textwidth]{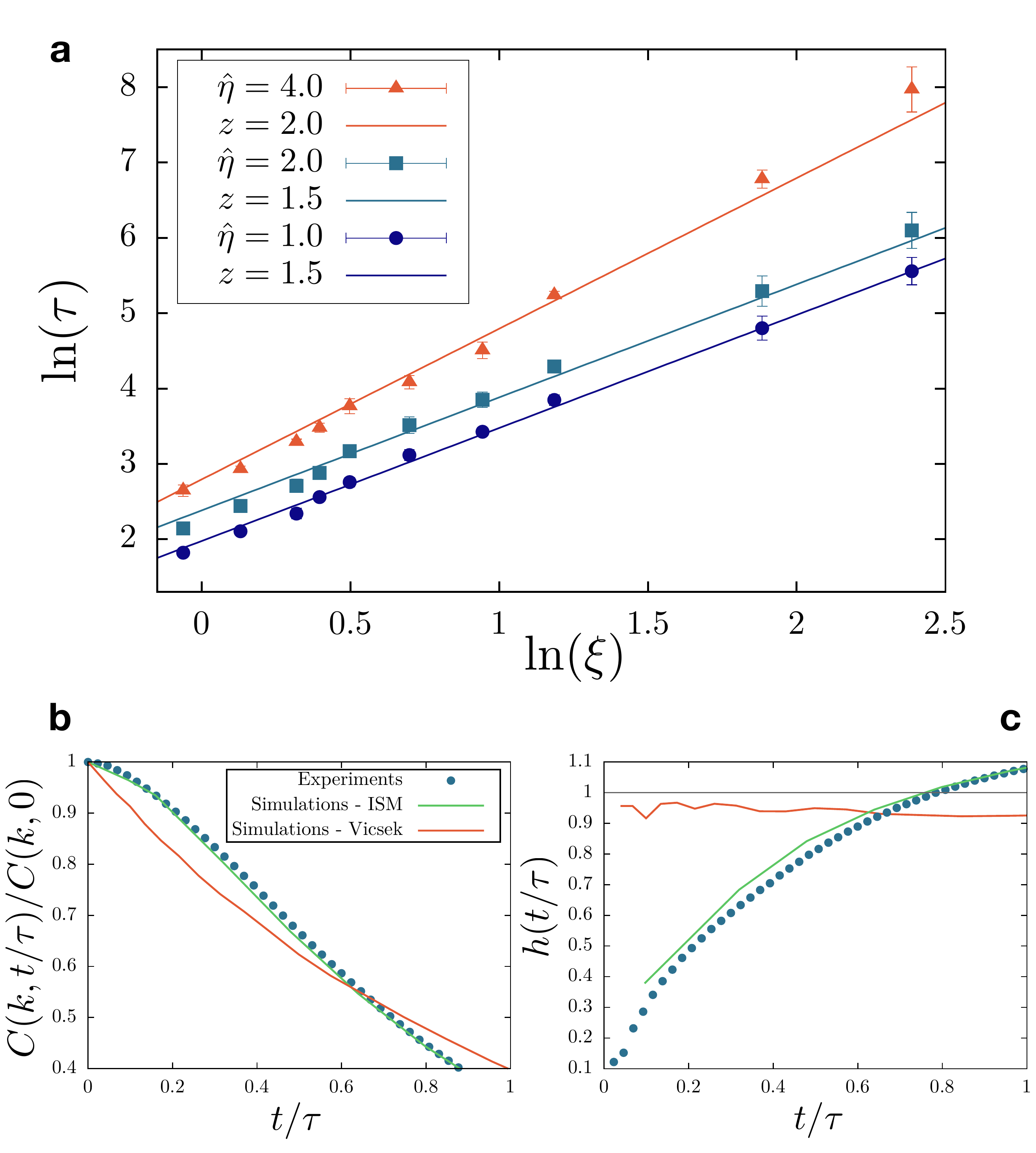}\\
\caption{{\bf Numerical simulations.} Top: Relaxation time vs correlation length in $d=3$, at various values of the friction $\hat\eta$. Lines are best fit to $z=3/2$ (blue - low $\hat\eta$) and $z=2$ (orange - large $\hat\eta$). Bottom, left: The normalized dynamical correlation function, $C(k,t)/C(k,0)$, at $k=1/\xi$. Right: The relaxation form factor, $h(t/\tau) \equiv \dot C(t/\tau)/C(t/\tau)$, goes to $1$ for exponential relaxation, while it goes to $0$ for inertial relaxation. Experimental data on swarms and data on the $3d$ Vicsek model close to the ordering transition, are from \cite{cavagna2017dynamic}.
}
\label{simu}
\end{figure}

To test our results we run numerical simulations of the microscopic ISM on a fixed lattice with periodic boundary conditions. The dynamical equations are \cite{cavagna+al_15},
\begin{align}
\frac{d \bm\psi_i}{d t} &=\frac{1}{\hat \chi}  \bm s_i \times \bm\psi_i  \ ,
\label{ISM1_micro} \\
\frac{d \bm s_i}{d t} &= - \frac{\hat\eta}{\hat\chi}  \bm s_i +  \bm\psi_i \times  \hat J  \sum_{j\in i} n_{ij} \bm\psi_j + \bm \psi_i \times \bm \zeta_i \ ,
\label{ISM2_micro}
\end{align}
where $\lvert\bm\psi_i\rvert^2=1$, the adjacency matrix $n_{ij}$ corresponds to a simple cubic lattice ($d=3$), $\hat J$ is the strength of the alignment interaction, and the hats distinguish microscopic from coarse-grained parameters.  Note that the microscopic model lacks a transport term $\hat \lambda\nabla^2 \bm s$, which originates when coarse-graining the spin.  
We compute the $k=0$ relaxation time $\tau$ in systems with linear sizes up to $L=20$ \cite{companion}. In order to observe the power-law crossover of (\ref{cross1}-\ref{cross2}) in a $\tau$ vs $\xi$ plot, one would need a span of $\xi$ of several orders of magnitudes, which is not possible with our maximum size
\footnote{Actually, this is a problem in general: three decades, $L=10^3$, is the {\it very} minimum needed to observe a power-law crossover. This gives $N=10^9$ in $d=3$, which is quite awful, especially considering that the dissipative relaxation time grows as $\tau \sim \xi^2 \sim L^2 \sim 10^6$.}.
Therefore, to test the predicted crossover we run simulations at different values of the effective friction: for small enough $\hat\eta$ we should have $\R0 > L$, so that the whole system is within the conservation scale and we expect to observe $\tau \sim \xi^{3/2}$ in the whole $\xi$ range; conversely, for large enough $\hat \eta$, we should have $\R0$ as small as the lattice spacing, giving $\tau \sim \xi^2$ at all $\xi$. This RG prediction is fully confirmed by numerical simulations (Fig.2a): the dynamical critical exponent $z$ crosses over from $3/2$ to $2$ by increasing the effective friction. Moreover, simulations show that the ISM dynamical correlation function in the critical regime has the same inertial form as real biological swarms, a significant improvement over the Vicsek model (Fig.2b,c). In particular, the relaxation form factor, $h(t/\tau) \equiv \dot C(t/\tau)/C(t/\tau)$  \cite{cavagna2017dynamic}, goes to $1$ for over-damped exponential relaxation (Vicsek model), while it goes to $0$ for the inertial ISM, exactly as in real swarms. 

The RG results bring us to an interesting conclusion: when a finite-size system has small dissipation it may become impossible to observe the effects of the violation of the conservation law on the critical exponents, simply because the whole system is smaller than the conservation length scale, $\R0$, and one inherits the same relaxation dynamics as the fully conservative RG fixed point. This observation leads us quite naturally to biological groups. The shape of the dynamical correlation functions in natural swarms provides experimental evidence that dissipation is small in these systems \cite{cavagna2017dynamic}. Because real swarms are of course finite-size systems, the present RG calculation suggests that dynamical scaling is ruled by the conservative fixed point, that is by the dynamical critical exponent $z=3/2$. Although this is still some way from the experimental value, $z\approx 1$, it is significantly better than the purely dissipative value $z\approx 2$ of the Vicsek model \cite{cavagna2017dynamic}. We conclude that a theory with the same social force as classic collective behaviour models, but where the force acts on the local velocities through a non-dissipative coupling, not only gives much more compelling correlation functions, but it also shifts the dynamical critical exponent $z$ in the right direction. Considering how difficult it is to change critical exponents in collective physical systems, we think this result is encouraging. Further studies will show whether going self-propelled  will bridge the gap between $3/2$ and $1$, or whether some other dynamical ingredients are required to match theory and experiments in natural swarms.

We thank Enzo Branchini, Erwin Frey and Luca Peliti for important discussions and suggestions, and Stefania Melillo for help with the experimental data of \cite{cavagna2017dynamic}.  
This work was supported by ERC Advanced Grant RG.BIO (contract n.785932) to AC, and ERANET-CRIB Grant to AC and TSG.  TSG was also supported by grants from CONICET, ANPCyT and UNLP (Argentina). 

\bibliographystyle{apsrev4-1}
\bibliography{general_cobbs_bibliography_file}

\end{document}